# The Distribution of Program Sizes and Its Implications: An Eclipse Case Study

Hongyu Zhang [*][†]

*School of Software, Tsinghua University, Beijing 100084, China.*

Hee Beng Kuan Tan

*School of Electrical & Electronic Engineering, Nanyang Technological University, Singapore 639798*

Michele Marchesi

*Department of Electrical and Electronic Engineering, University of Cagliari, Cagliari 09123, Italy*

**SUMMARY.**

**A large software system is often composed of many inter-related programs of different sizes. Using the public Eclipse dataset, we replicate our previous study on the distribution of program sizes. Our results confirm that the program sizes follow the lognormal distribution. We also investigate the implications of the program size distribution on size estimation and quality predication. We find that the nature of size distribution can be used to estimate the size of a large Java system. We also find that a small percentage of largest programs account for a large percentage of defects, and the number of defects across programs follows the Weibull distribution when the programs are ranked by their sizes. Our results show that the distribution of program sizes is an important property for understanding large and complex software systems.**

## 1. INTRODUCTION

Contemporary software development, such as object-oriented paradigm, follows the principle of modularization. In contrasting to "monolithic" software development, modularization decomposes a large solution space into a set of smaller and manageable programs, which could be separately developed and then be composed to form an executable software system. A well-designed program encapsulates certain information, separates certain "concerns" and communicates with other programs via interfaces.

Size is an important static attribute of software, which is often counted using the Lines of Code (LOC) metric. A large, complex software system is usually composed of many inter-related programs of different sizes. It is desirable to understand the distribution of the sizes among programs in order to better understand the complexity of a software system. In recent years, researchers have applied complex system theory (and in general theoretical physics) to study various properties of a large software system including size. It is found that the distribution of program size follows the lognormal distribution (Concas et al., 2007; Zhang and Tan, 2007).

---

[*]   Correspondence to: Hongyu Zhang, School of Software, Tsinghua University, Beijing 100084, China.
[†]   E-mail: hongyu@tsinghua.edu.cn

In this paper, we use the public Eclipse dataset (Zimmermann et al., 2007) to replicate the study reported in our previous paper (Zhang and Tan, 2007). Eclipse is a widely used integrated development platform for creating Java, C++ and web applications. The public Eclipse dataset contain measurement and defect data for Eclipse versions 2.0, 2.1 and 3.0. These three Eclipse systems can be considered as large software systems --- they contain in average 1030KLOC and 8403 programs. We observe the *small program phenomenon* through the analysis of the dataset: that most programs are small while only a small percentage of programs are very large. Furthermore, the distributions of program sizes in these three Eclipse systems are indeed lognormal distributions.

We also investigate the implications of size distributions on size estimation. Having understood the lognormal nature of size distribution, we are able to estimate the total size of a large Java system, and to estimate the number of programs within a certain size range. The estimation is based on a general size distribution model that is obtained from a corpus of Java systems. In this paper, we also explore the implications of size distributions on defects. We find that a small percentage of largest programs account for a large percentage of defects, and the number of defects across programs follows the Weibull distribution when the programs are ranked by their sizes.

We believe our study reveals the regularity that emerges from large-scale software construction. Our results show that the distribution of program sizes is an important property for understanding large and complex software systems.

## 2. THE DISTRIBUTION OF PROGRAM SIZES

### 2.1 The Eclipse size data

We firstly perform descriptive analysis of program sizes for the studied Eclipse systems. The program size is measured using the Lines of Code (LOC) metric, which counts each physical source line of code in a program, excluding blank lines and comments. Although some authors have pointed out the deficiencies of LOC, it is still the most commonly used size measure in practices because of its simplicity.

Table 1 shows the descriptive statistics of program size data, which describes the central tendency and dispersion of the sizes. We find that for all Eclipse versions, the mode values of size (the value of the most commonly occurring item) are close to the min (minimum) values and are smaller than the median values (the value of the middle-ranked item). The median values are smaller than the mean values (the average value) and the max (maximum) values are much greater than other values. Therefore the descriptive statistics indicate that the distribution of program sizes is a highly skewed one. This is confirmed by Figure 1, where the programs are ranked by their size (from the largest to smallest). Clearly all curves in Figure 1 exhibit the "long tail" behavior, showing that most programs are small while only a few programs are very large. As an example, the data shows that 38.03% of the Eclipse 3.0 programs are smaller than 32 LOC, 56.42% of the Eclipse 3.0 programs are smaller than 64 LOC. Although many programs are small, there are still a small number of very large programs: about 4.39% of Eclipse 3.0 programs are larger than 512 LOC and 1.13% of Eclipse 3.0 programs are larger than 1024 LOC. We call this phenomenon the *small program phenomenon*[1].

**Table 1:** The descriptive statistics of the Eclipse program sizes in LOC.

| Version | #Programs | Min | Median | Max | Mode | Mean | Std Dev. |
|---|---|---|---|---|---|---|---|
| 2.0 | 6729 | 3 | 51 | 5207 | 7 | 118.43 | 219.98 |
| 2.1 | 7888 | 3 | 54 | 5228 | 7 | 125.20 | 233.79 |
| 3.0 | 10593 | 3 | 51 | 4886 | 5 | 123.28 | 233.49 |

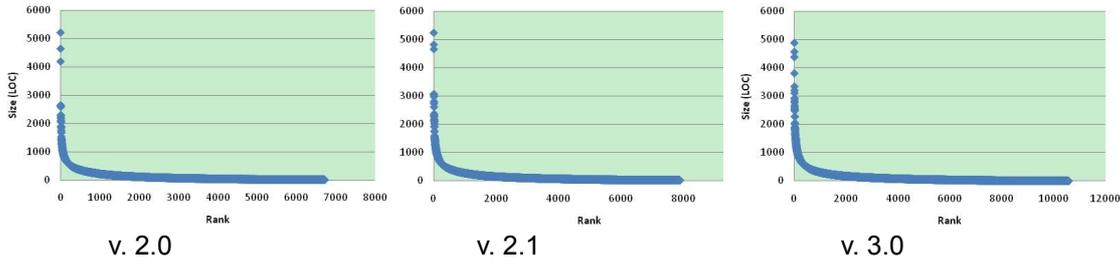

v. 2.0     v. 2.1     v. 3.0

**Fig. 1.** The distribution of program sizes in Eclipse.

### 2.2 The lognormal distribution of program sizes

A random variable X has a lognormal distribution if the random variable $Y = ln(X)$ is normally distributed, where $ln(X)$ is the natural logarithm of X. The lognormal distribution has probability density function:

$$f(x) = \frac{1}{\sigma x \sqrt{2\pi}} \exp\left(\frac{-(\ln x - \mu)^2}{2\sigma^2}\right), \quad x>0 \; \sigma>0 \qquad (1)$$

where μ and σ are the mean and standard deviation of the variable's logarithm. μ is also called the scale parameter and σ the shape parameter.

In our earlier work (Zhang and Tan, 2007), we analyzed a corpus of Java software systems and found that their program sizes all follow the lognormal distribution. The corpus contains 18 randomly collected, large-scale open source Java systems. These systems have in average 2770 programs and 339K line of code. For these systems, the sample means of lognormal parameters μ and σ are 3.8277 and 1.3472, respectively.

Using the Eclipse dataset, we replicate the study described in our earlier work. To statistically compare the goodness-of-fit of a fit, we compute the coefficient of determination ($R^2$) and the Standard Error of Estimate ($S_e$). The $R^2$ statistic measures the percentage of variations that can be explained by the model. Its value is between 0 and 1, with higher value indicating a better fit. $S_e$ is a measure of the absolute prediction error and is computed as:

$$S_e = \sqrt{\frac{\sum (y - y')^2}{n-2}}, \qquad (2)$$

---
[1] As in Java, a program (.java file) can only include one public class. The complexity of private and inner classes can be counted into the associated public classes. Therefore, sometimes we also call the *small program phenomenon* "*small class phenomenon*".

where y and y' are the actual and predicted values, respectively. Larger $S_e$ indicates the larger prediction error.

We find that the lognormal distribution can model fairly well the distribution of program sizes in all three Eclipse versions. Table 2 shows the obtained lognormal parameters and the accuracy measures. The $R^2$ values range from 0.9976 to 0.9979, and the $S_e$ values range from 0.0161 to 0.0171. These results confirm that the program size distribution can be indeed modeled by lognormal distribution.

**Table 2.** The lognormal parameters for Eclipse systems

| Version | μ | σ | $R^2$ | $S_e$ |
|---|---|---|---|---|
| 2.0 | 3.9006 | 1.3451 | 0.9979 | 0.0161 |
| 2.1 | 3.9383 | 1.3621 | 0.9978 | 0.0166 |
| 3.0 | 3.9006 | 1.3744 | 0.9976 | 0.0171 |

## 3. THE IMPLICATION OF SIZE DISTRIBUTION ON SIZE ESTIMATION

### 3.1 Estimating the total size of a large Java system

Having understood the lognormal nature of size distribution, we can then estimate sizes of large Java systems as follows:

$$Size = N * \int \frac{x}{\sigma x \sqrt{2\pi}} \exp\left(\frac{-(\ln x - \mu)^2}{2\sigma^2}\right) dx \qquad (3)$$

where $N$ is the number of programs. An analytic solution for this is:

$$Size = N * \exp(\mu + \sigma^2/2) \qquad (4)$$

The Equation (4) shows that we can estimate software size solely based on the number of programs if the lognormal parameters μ and σ are known.

To do this, we use the sample means of μ and σ (3.8277 and 1.3472) to estimate the population means. The sample means are obtained from our earlier study of 18 Java systems (Zhang and Tan, 2007). The Equation (4) then becomes:

$$Size = N * \exp(3.8277 + (1.3472)^2/2) = N * 113.88 \qquad (5)$$

The Equation (5) reveals the remarkably simple relationship between the size of a large Java software system and the number of programs the system has – that the size is a product of $N$ (the number of programs) and 113.88. The constant 113.88 can be seen as the expected average size of Java programs.

**Table 3.** Size estimation based on the nature of program size distribution

| Version | Programs (N) | Actual Size (LOC) | Estimated Size (LOC) | MRE | MRE≤25% |
|---|---|---|---|---|---|
| 2.0 | 6729 | 796941 | 766299 | 3.85% | Yes |
| 2.1 | 7888 | 987603 | 898285.4 | 9.04% | Yes |
| 3.0 | 10593 | 1305908 | 1206331 | 7.63% | Yes |

We now use the Eclipse data to evaluate the Equation (5). The results are shown in Table 3. We use the MRE (Magnitude of Relative Error) metric to measure the estimation accuracy.

MRE is defined as *100% * | N – N^|/N*, where *N* and *N^* are the actual value and its estimate, respectively. The commonly acceptable criterion is MRE≤0.25 (the smaller the value the better the estimation). In Table 3, the MRE values for the three Eclipse versions range from 3.85% to 9.04%. The estimations are satisfactory. This result shows that it is possible to estimate the total size of a large Java system if its size distribution follows the lognormal distribution. Currently, the scope of our experiments is still limited. It would be interesting to find out if such estimation can be applied to other Java systems and, if not, why.

### 3.2 Estimating the number of programs within a size range

Based on the lognormal nature of the distribution, we can also estimate the number of programs within a size interval. For example, we can estimate the number of programs between a given size range [x1, x2] as follows:

$$N_{interval} = N * P(x1 \leq x \leq x2) = N * (P(x \leq x2) - P(x \leq x1)) \qquad (6)$$

where *N* is the total number of the programs in a system, *P* is the cumulative density function (CDF) of the lognormal distribution. Again, we can use the sample means of lognormal parameters $\mu$ and $\sigma$ (3.8277 and 1.3472) to characterize a general model of *P*.

As an example, for Eclipse 3.0 we know that it contains 10593 programs. Suppose we want to know the approximate number of large programs sized from 1024 LOC to 2048 LOC in Eclipse 3.0, using the Equation (6) we get the following estimate:

$$N_{interval} = N * P(x1 \leq x \leq x2) = 10593 * (P(x \leq 2048) - P(x \leq 1024))$$
$$= 10593 * (0.9976 - 0.9894) = 87 \qquad (7)$$

In Eclipse 3.0, the actual number of programs sized in the range of [1024, 2048] is 99. The MRE is only 12.12%, which is within the criterion of acceptability.

Using the Eclipse data, we experiment with different size ranges. The results are shown in Table 4. All MRE values except one are less than 25%. The exceptional case (the range [1024, 2048] for Eclipse 2.0) is just slightly above the criterion, and the absolute difference is low. Therefore, we conclude that it is possible to estimate the number of programs within a size range based on the general lognormal size distribution model.

**Table 4.** Estimating the number of Eclipse programs within a size range

| Version | Size Range | Actual #programs | Estimated #programs | MRE | MRE<=25% |
|---------|------------|------------------|---------------------|--------|----------|
| 2.0     | [3, 64]    | 3814             | 3874                | 1.57%  | Yes      |
|         | [65, 256]  | 2115             | 2000                | 5.42%  | Yes      |
|         | [257, 1024]| 743              | 606                 | 18.44% | Yes      |
|         | [1025, 2048]| 43              | 55                  | 28.03% | No       |
| 2.1     | [3, 64]    | 4362             | 4541                | 4.11%  | Yes      |
|         | [65, 256]  | 2528             | 2345                | 7.24%  | Yes      |
|         | [257, 1024]| 914              | 710                 | 22.28% | Yes      |
|         | [1025, 2048]| 64              | 65                  | 0.84%  | Yes      |
| 3.0     | [3, 64]    | 5977             | 6098                | 2.03%  | Yes      |
|         | [65, 256]  | 3348             | 3149                | 5.94%  | Yes      |
|         | [257, 1024]| 1148             | 954                 | 16.90% | Yes      |
|         | [1025, 2048]| 99              | 87                  | 12.12% | Yes      |

## 4. The IMPLICATION OF SIZE DISTRIBUTION ON DEFECTS

It is widely believed that some internal properties of software, such as size, have correlations with software quality (Fenton and Pfleeger, 1997). A number of previous studies investigated the impact of size on program complexity (Li and Cheung, 1987) and the number of faults (El Emam et al., 2002). Many defect prediction models have been proposed based on the measurement of internal program properties including LOC (e.g., Koru and Liu, 2005; Menzies et al., 2007; Zhang, Zhang and Gu, 2007). In general, larger programs are more likely to contain faults than smaller modules, and are more likely to have high complexity.

We believe that the skewed distribution of program size also implies that the distribution of defects across programs is skewed. In this study, we use the public Eclipse defect data to explore the implication of size distribution on defects. The Eclipse defect data is collected from Eclipse's bug databases and version achieves (Zimmermann et al., 2007). There are two kinds of defects: pre-release defects (defects reported in the last six months before release) and post-release defects (defects reported in the first six months after release).

For each Eclipse version, we rank the programs according to their size (from the largest LOC to the smallest LOC), and then calculate the cumulative percentage of defects. We find that the distribution of defects is highly skewed --- that a small number of largest programs accounts for a large proportion of the defects. For example, the top 10% of the largest Eclipse 3.0 programs are responsible for 46.28% pre-release defects and 44.05% post-release defects. The top 20% of the largest Eclipse 3.0 programs are responsible for 62.29% pre-release defects and 60.62% post-release defects. This is also illustrated in Figure 2, which shows the relationship between the cumulative percentage of modules and the cumulative percentage of defects in Eclipse 3.0. Similar phenomenon is observed in other Eclipse versions (Table 5). The results confirm that, by simply ranking the programs according to their sizes, a large number of defects can be located. The results also suggest that, in software quality assurance practices (such as inspection, formal review meeting, and testing), if initial efforts are centered on the largest programs, a large percentage of defects could be identified earlier.

**Table 5.** Defects contained by the top *x*% largest programs in Eclipse

|  | Version | Top 5% | Top 10% | Top 15% | Top 20% | Top 25% |
|---|---|---|---|---|---|---|
| Pre-release defects | 2.0 | 24.57% | 37.01% | 46.99% | 53.48% | 60.33% |
|  | 2.1 | 28.82% | 43.46% | 53.97% | 61.01% | 68.20% |
|  | 3.0 | 32.98% | 46.28% | 55.05% | 62.29% | 68.93% |
| Post-release defects | 2.0 | 34.16% | 46.87% | 55.73% | 61.88% | 67.85% |
|  | 2.1 | 28.09% | 40.52% | 47.72% | 54.31% | 60.49% |
|  | 3.0 | 29.97% | 44.05% | 52.41% | 60.62% | 67.53% |

Further analysis shows that the distributions shown in Figure 2 are actually Weibull distribution. Weibull distribution, developed by the physicist Waloddi Weibull, is one of the most widely used probability distributions in the reliability engineering discipline (Ramakumar, 1993). The CDF (cumulative density function) of the Weibull distribution can be formally defined as:

$$P(x) = 1 - \exp\left(-\left(\frac{x}{\gamma}\right)^{\beta}\right) \quad (\gamma > 0, \beta > 0) \tag{8}$$

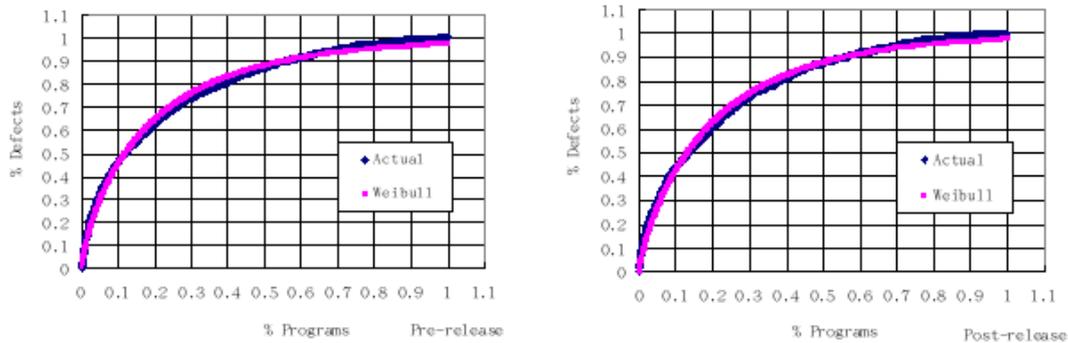

**Fig. 2:** The Weibull distribution of defects when programs are ordered
by LOC (Eclipse 3.0)

Using statistical packages such as the SPSS, we are able to perform non-linear regression analysis and derive the parameters for each distribution. Figure 2 also shows the fitted curves of Weibull distribution for both pre-release defects and post-release defects in Eclipse 3.0. Clearly the Weibull distribution fits the actual data well. Table 6 below gives the Weibull parameters and the accuracy measures for all Eclipse versions we studied. The $R^2$ value ranges from 0.986 to 0.995, and the $S_e$ values range from 0.017 to 0.026. These results confirm that defects follow the Weibull distribution when programs are ordered by LOC.

**Table 6.** The Weibull distribution of Eclipse defects when programs are ordered by LOC

|  | **Version** | $\gamma$ | $\beta$ | $R^2$ | $S_e$ |
|---|---|---|---|---|---|
| Pre-release defects | 2.0 | 0.259 | 0.897 | 0.991 | 0.023 |
|  | 2.1 | 0.207 | 0.830 | 0.995 | 0.017 |
|  | 3.0 | 0.193 | 0.780 | 0.992 | 0.019 |
| Post-release defects | 2.0 | 0.190 | 0.811 | 0.986 | 0.026 |
|  | 2.1 | 0.242 | 0.853 | 0.988 | 0.026 |
|  | 3.0 | 0.203 | 0.827 | 0.993 | 0.019 |

## 5. RELATED WORK

Some other researchers also observed the existence of a large proportion of small programs in systems they studied. For example, Wilde et al. (1993) studied maintainability of three object-oriented systems written in C++ and Smalltalk. Two systems were developed in Bell Communication Research (Bellcore), one system was a PC Smalltalk environment that has an extensive class library. They found that many methods in the classes have small sizes --- in all three systems, half or more of the methods are fewer than four Smalltalk lines or two C++ statements. Chidamber and Kemerer (1994), during their study of design complexity of two C++ and Smalltalk systems, found that "most classes tend to have a small number of methods (0 to l0), while a few outliers declare a large number of them". We believe that the study of program size distribution is important for understanding various characteristics of a large software system. Therefore we dedicate our study to the issues relating to program size distribution. We discover the "small program phenomenon" and also formally describe it using the lognormal distribution.

In recent years, some researchers have started to apply complex system theory (and in general theoretical physics) to the measurement of a large software system. For example, Baxter et al. (2006) collected a corpus of Java software and analyzed the structural attributes of the programs (such as Number of Fields, Number of Constructors, etc). They discovered that the distribution of some attributes follow power-law while others do not. Concas et al. (2007) found the Pareto and lognormal distributions in many software properties such as the number of method calls, in-degree and out-degree of class dependency graph, etc. Our work is similar in spirit to these work as we all try to discover the underlying regularity behind software construction. We focus on the distribution of program sizes and also discuss its implications on size estimation and quality prediction.

Many researchers have studied the relationship between size and defects. Fenton and Ohlsson (2000) used a diagram (called *Alberg* diagram) to display the accumulated percentage of the number of defects when programs are ordered with respect to LOC for a telecommunication system. They concluded that LOC is quite good at ranking the top 20% of the defect-prone programs. They termed it the "ranking ability" of LOC. However, they did not elaborate their findings, nor provided any detailed statistical analysis. Ostrand et al. (2005) studied the "ranking ability" of LOC for an inventory system and a provisioning system. They found that 20% of largest files contain 73% and 74% of the defects for the two systems. Andersson and Runeson (2007) replicated the study of Fenton and Ohlsson. Their results did not conclusively support the LOC's ranking ability on defect-proneness. In our prior study, we discovered that the distribution of defects over modules (when modules are ranked according to the number of defects) can be better modeled as the Weibull distribution (Zhang, 2007). In this paper, we study the ranking ability of LOC using the public Eclipse dataset. We discover the Weibull distribution of defects when programs are ranked by LOC.

## 6. DISCUSSIONS AND CONCLUSIONS

A large software system is usually composed of many programs of different sizes. In this paper, we have used the public Eclipse dataset to replicate our previous study on the program size distribution. The results confirm the *small program phenomenon* and the lognormal distribution of program sizes. We find that the nature of size distribution can be used to estimate the total size of a large Java system, and to estimate the number of programs within a certain size range. We also find that a small percentage of largest programs account for a large percentage of defects, and the number of defects across programs follows the Weibull distribution when the programs are ranked by their sizes.

Our results show that we can achieve better understanding of a large-scale software system if we could understand the distribution of its program sizes. By using the public Eclipse datasets, we hope that other researchers could replicate our experiments and advance this research.

There are several threats to validity of our results. Until now, all our experiments are based on a small sample of randomly chosen large-scale Java systems. We derived the lognormal distribution of programs sizes from the analysis of these systems. The size estimations are also based on the sample means. If this sample does not represent the true characteristics of the population, our results could be biased.

In future, we will further evaluate the distribution of program sizes for a wide variety of large-scale software systems, and to verify the validity of the proposed size estimation methods. We will also explore the generation mechanism of the lognormal distribution. It is found that program size has correlations with development and maintenance efforts (Boehm, 1981). In general, programs with larger size are more likely to require more development and maintenance effort. Understanding the implications of size distribution on development and maintenance efforts is also an important future work. We hope our work could be helpful for the theory and practices of computer programming.

## Acknowledgments

This research is supported by the Chinese NSF grants 60703060, 90718022, and the state 863 project 2007AA01Z480.